%% file: p4-sfu.tex
\documentclass[sigconf,nonacm,authorversion]{acmart}

\renewcommand\footnotetextcopyrightpermission[1]{} 

\usepackage{tikz}
\usepackage{amsmath}
\usepackage{makecell}
\usepackage{xcolor}
\usepackage{caption}
\usepackage{balance}
\usepackage{subcaption}
\usepackage{enumitem}
\usepackage{titlesec}

\newcommand{\squishlist}{
   \begin{list}{$\bullet$}
    { \setlength{\itemsep}{0pt}      \setlength{\parsep}{3pt}
      \setlength{\topsep}{3pt}       \setlength{\partopsep}{0pt}
      \setlength{\leftmargin}{1.2em} \setlength{\labelwidth}{1em}
      \setlength{\labelsep}{0.5em} } }
\newcommand{\squishend}{
    \end{list}  }

\titlespacing{\section}{0pt}{10pt}{2pt}
\titlespacing{\subsection}{0pt}{5pt}{2pt}

\newcommand{\boldpara}[1]{\noindent \textbf{#1}}

\newcommand{\jrex}[1]{}
\newcommand{\ravi}[1]{}
\newcommand{\joon}[1]{}
\newcommand{\oliver}[1]{}
\newcommand{\sata}[1]{}

\newcommand{\sys}{Scallop}

\renewcommand\footnotetextcopyrightpermission[1]{} 
\setcopyright{none}
\hypersetup{hidelinks}

\microtypecontext{spacing=nonfrench}

\acmDOI{}

\acmISBN{}

\acmConference[Submitted for review]{}
\acmYear{2025}
\copyrightyear{}

\acmPrice{}

\begin{document}

\date{}

\title{Scalable Video Conferencing Using SDN Principles}



\author{Oliver Michel}
\affiliation{Princeton University\country{}}
\email{omichel@princeton.edu}

\author{Satadal Sengupta}
\affiliation{Princeton University\country{}}
\email{satadals@princeton.edu}

\author{Hyojoon Kim}
\affiliation{University of Virginia\country{}}
\email{tcr5zr@virginia.edu}

\author{Ravi Netravali}
\affiliation{Princeton University\country{}}
\email{rnetravali@princeton.edu}

\author{Jennifer Rexford}
\affiliation{Princeton University\country{}}
\email{jrex@princeton.edu}

\begin{abstract}
    \noindent
    Video-conferencing applications face an unwavering surge in traffic, stressing their underlying infrastructure in unprecedented ways.
    This paper rethinks the key building block for conferencing infrastructures --- selective forwarding units (SFUs).
    SFUs relay and adapt media streams between participants and, today, run in software on general-purpose servers.
    Our main insight, discerned from dissecting the operation of production SFU servers, is that SFUs largely mimic traditional packet-processing operations such as dropping and forwarding.
    Guided by this, we present Scallop, an SDN-inspired SFU that decouples video-conferencing applications into a hardware-based data plane for latency-sensitive and frequent media operations, and a software control plane for the (infrequent) remaining tasks, such as analyzing feedback signals. Our Tofino-based implementation fully supports WebRTC and delivers 7-210$\times$ improved scaling over a 32-core commodity server, while reaping performance improvements by cutting forwarding-induced latency by 26$\times$.
\end{abstract}

\maketitle

\pagestyle{plain}

\input{sec/1-intro}

\noindent
\textbf{Ethics.}~%
The campus traces and the Zoom API data used in this study were anonymized with a one-way hash.
All data were inspected and sanitized by an authorized network operator to remove all personal data before being accessed by researchers.
In the packet traces, the media payloads have been removed as well before researchers gained access.
This study has been conducted with necessary approvals from our institution, including its Institutional Review Board (IRB).

\input{sec/2-sfu-scaling}
\input{sec/3-scallop}

\input{sec/4-control}
\input{sec/5-data-plane}

\input{sec/6-eval}

\input{sec/7-discussion}

\input{sec/8-related}
\input{sec/9-conclusion}

\balance{}

\bibliographystyle{ACM-Reference-Format}
\bibliography{p4-sfu}

\clearpage
\nobalance{}

\appendix
\input{sec/a5-discussion}
\input{sec/a2-zoom-api}
\input{sec/a3-zoom-passive}
\input{sec/a4-zoom-simulcast}

\input{sec/a6-rtp-extension-parsing}

\section{Tofino Resource Utilization}
\label{apx:tofino-resource}

\noindent
Table~\ref{tab:tofino-usage} summarizes the resource utilization of the Tofino as described in Section~\ref{sec:data-plane}.

\begin{table}[h]
    \centering
    \footnotesize
    \begin{tabular}{p{1.9cm}|p{1.4cm}|p{1.7cm}|p{1.7cm}}
        \textbf{Resource type} & \textbf{Scaling behavior} & \textbf{Usage under peak campus load (avg.)} & \textbf{under max util. (avg.)} \\
        \hline
        Parsing depth & Fixed & Ing. 27, Eg. 7 & = \\
        No. of stages & Fixed & Ing. 7, Eg. 5 & = \\
        PHV containers & Fixed & 17.9\% & = \\
        Exact xbars & Fixed & 5.66\% & = \\
        Ternary xbars & Fixed & 2.52\% & = \\
        Hash bits & Fixed & 4.62\% & = \\
        Hash dist. units & Fixed & 6.94\% & = \\
        VLIW instr. & Fixed & 7.29\% & = \\
        Logical table ID & Fixed & 21.87\% & = \\
        SRAM & Fixed & 6.77\% & = \\
        TCAM & Fixed & 1.38\% & = \\
        \hline
        Egress Tput. & Quadratic &  1.2 Gb/s & 197 Gb/s \\
        \hline
    \end{tabular}
    \vspace{0.2cm}
    \caption{Resource usage of the Tofino data plane.}
    \label{tab:tofino-usage}
        \vspace{-0.3cm}
\end{table}

\end{document}

%% file: sec/1-intro.tex

\section{Introduction}%
\label{sec:intro}

\noindent
Video-conferencing applications (VCAs) such as Google Meet and Zoom have become essential tools for remote work, education, healthcare, and social interactions. The past decade has seen substantial efforts to improve these applications, including via 
more efficient video codecs~\cite{zhang:loki, fouladi:salsify, schwarz:svc-overview}, rate-adaptation algorithms~\cite{carlucci:google-cc, decicco:cc-for-webrtc, engelbart:cc-for-real-time-over-quic, rfc8698}, and measurement studies on user experience~\cite{bronzino:inferring-video-quality, michel:passive-measurement-zoom, chang:can-you-see-me, macmillan:perf-video-conferencing}. Though effective, prior work has primarily focused on end users, with the scaling challenges that VCA \emph{operators} face to support exploding traffic rates~\cite{feldmann:lockdown-effect, video-conf-market} being far less explored.

At the core of VCA infrastructure are selective forwarding units (SFUs)~\cite{gouaillard:webrtc-sfu-load-test,jitsi-videobridge,mediasoup,lopez:kurento}. These servers are tasked not only with relaying video and audio streams among meeting participants, but also with monitoring and adapting those media signals to match the time-varying network and compute capabilities of users. Unfortunately, their deployments on general-purpose servers -- the status quo today -- makes scaling very difficult, particularly given several fundamental properties of VCAs (\S\ref{sec:sfu-scaling}):
\squishlist

\item The workload of an SFU is \emph{hard to predict and can change rapidly}. Beyond diurnal variation, the number of streams that an SFU must handle grows quadratically with the number of participants in a meeting, i.e., even a single new participant in a meeting introduces substantial load since their media streams must be relayed to all other participants, and they must receive all media streams from all other participants.

\item And yet, the replication and forwarding that SFUs perform on media packets is on the \emph{critical path of user interactions}. At hundreds of packets/sec/stream, operating-system delays for software packet processing (e.g., scheduling, context switches, interrupts, socket-buffer copying) can lead to significant user-perceived jitter and latency, especially in the face of under-provisioned resources.

\squishend
As a result, VCA operators are left with two options today: massively over-provision SFU server infrastructure to handle peak loads, which is costly and wasteful, or (reactively) auto-scale those resources using traditional mechanisms~\cite{verma:borg, roy:efficient-autoscaling} which risks harming QoE for users.

In this paper, we forego ephemeral SFU scaling enhancements (e.g., improved software packet processing or provisioning mechanisms) in favor of a fundamental rethink of VCA infrastructure that can support long-term traffic forecasts. Guided by our detailed study of production SFUs and real campus VCA traces (\S\ref{sec:sfus-as-packet-processors}), our key insight is that: despite the large semantic gap, the forwarding and adapting of media signals at SFUs -- the most frequent tasks that account for the lion's share of scaling overheads -- is \textit{strikingly similar to traditional packet-processing tasks}. Indeed, relaying and adapting media signals in today's SFUs can be distilled down to primitives such as packet replication and selective dropping that network hardware is optimized for.

Fueled by this insight, we present \textbf{Scallop} (a wordplay for ``scale up''),
a new hardware/software co-designed SFU that is built atop the WebRTC standard.
Drawing inspiration from SDN principles, Scallop decouples SFUs into (1) an efficient data plane that adapts and relays high-volume media streams using line-rate network switching hardware, and (2) a two-tier software control plane that handles the remaining infrequent tasks, e.g., session management, periodic feedback handling, and signaling.
The potential benefits are significant, with promises of 10-100$\times$ improved packet-processing performance over general-purpose servers at similar cost with fixed per-packet delays to eliminate SFU-induced jitter.


Yet, decoupling SFUs in this manner requires a rethink of their operation. Owing to the peer-to-peer (P2P) design of video-conferencing protocols like WebRTC, SFUs traditionally operate as \emph{split proxies} that terminate and spawn new client connections. However, this design would burden the SFU with tasks typically handled by end hosts and unsuited for network hardware (e.g., packet de/re-encryption), ultimately increasing control plane overheads and reducing scalability. Instead, we aim to run SFUs as \emph{true proxies} by capitalizing on the observation that most of these functions already run individually at clients and need not be replicated at SFUs. Doing so, while remaining faithful to existing protocols and SFU behavior poses three main challenges.

First, to reduce the event rate processed in the control plane, we use receiver-driven feedback in which \emph{receiving} clients continually handle bandwidth estimation and send only \emph{periodic} feedback reports to the SFU.
However, doing so naively in Scallop's proxy architecture would result in feedback signals from each receiver conflating information about multiple senders; this, in turn, would cause the transmission bitrate for each sender to ultimately converge to the lowest-bandwidth receiver, unnecessarily harming quality for others. To solve this problem, Scallop introduces new filtering and forwarding strategies in the data plane for relaying feedback that create the illusion of P2P sessions from each client's perspective, thereby preserving the ability to perform correct rate adaptation.

Second, \sys{}'s data plane must implement SFU-style media replication using hardware \emph{replication engines} primarily designed for traditional multicast. However, unlike IP multicast where all receivers get identical packet copies, P2P conferencing protocols require each SFU replica to be uniquely addressed per receiver. Further, hardware replication engines also have limited resources (e.g., a fixed number of multicast trees) and restricted flexibility for rate adaptation, creating potential bottlenecks. To address this, we design three specialized multicast tree structures optimized for different rate adaptation modes and dynamically migrate meetings between them, ensuring seamless transitions and maximizing hardware bandwidth utilization.

Third, rate adaptation requires dropping packets that contain media frames, but this creates sequence gaps that WebRTC receivers (which assume a P2P architecture) interpret as network losses, triggering erroneous retransmissions. To maintain the P2P illusion, the SFU must rewrite sequence numbers to mask intentional gaps in the stream. However, such rewriting is inherently difficult, especially in the presence of network-induced loss and reordering -- even software implementations cannot do this perfectly. 
Our experiments reveal that leaving extra gaps, however, is preferable to masking legitimate ones: missing sequence numbers trigger packet retransmissions, while incorrect rewrites break the decoder’s state, leading to a permanent freeze that can only be recovered from through a new key frame.
Based on this finding, we design a hardware-friendly sequence-number rewriting heuristic that minimizes retransmissions while preserving stream continuity, even under high loss or reordering.

We implemented Scallop using P4 on a 12.8 Tbit/s Intel Tofino2 ASIC. In experiments replaying campus-scale Zoom traces, Scallop handles 96.5\% of all packets and 99.7\% of bytes entirely in the hardware-based data plane. This enables Scallop to support up to 128,000 concurrent meetings on a single switch, a 7-210$\times$ improvement over a 32-core commodity server running existing SFUs~\cite{mediasoup}. Further, Scallop reduces the latency introduced by SFUs by a factor of 26.8, improving QoE for all participants. We will publicly release our Scallop implementation post publication.


%% file: sec/2-sfu-scaling.tex

\section{SFU Scaling Challenges}%
\label{sec:sfu-scaling}

\begin{figure}[t]
    \centering
    \includegraphics[width=0.9\columnwidth]{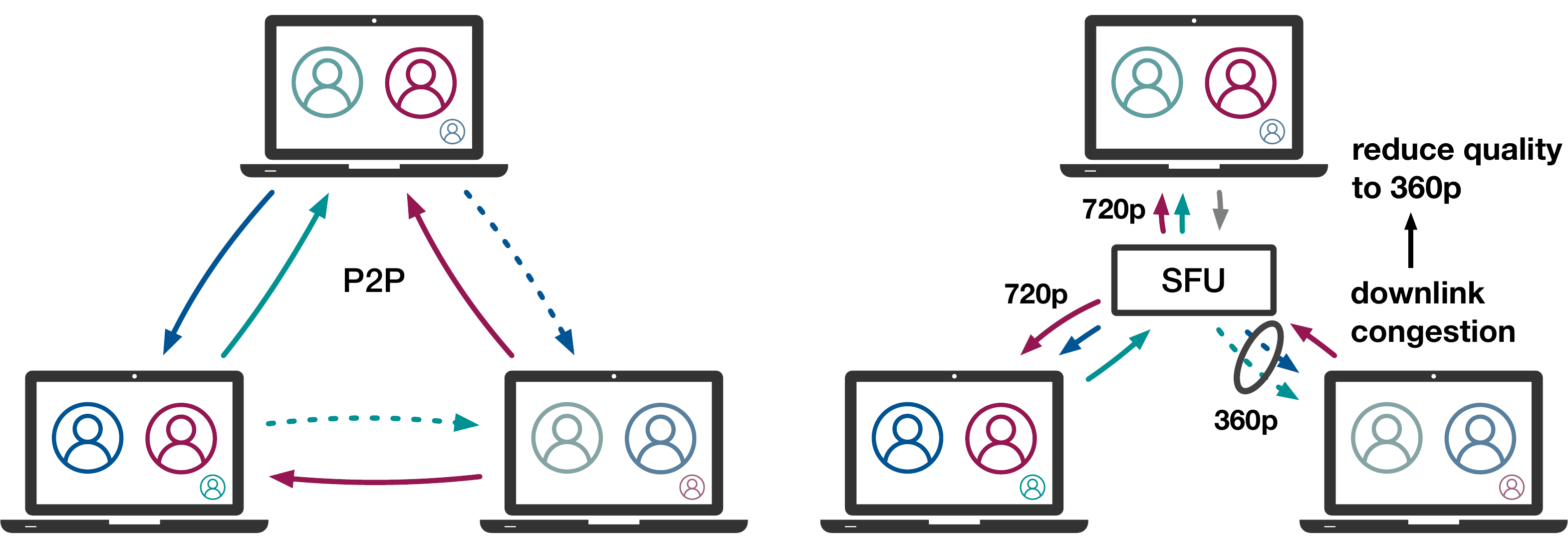}
    \caption{VCA Architectures: P2P vs. SFU.}
    \vspace{-6pt}
    \label{fig:call-arch}
\end{figure}

\begin{figure*}[t]
    \centering
    \begin{minipage}[t]{0.31\textwidth}
        \centering 
        \includegraphics[width=0.85\columnwidth]{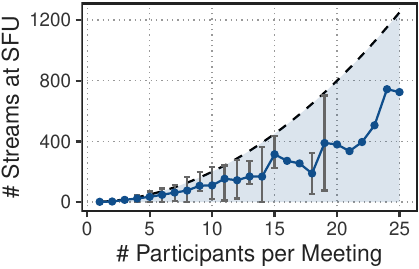}
        \caption{Number of media streams per meeting in campus trace.}
        \label{fig:n-sfu-streams}
    \end{minipage}
    \hspace{0.4cm}
    \begin{minipage}[t]{0.31\textwidth}
        \centering 
        \includegraphics[width=0.85\columnwidth]{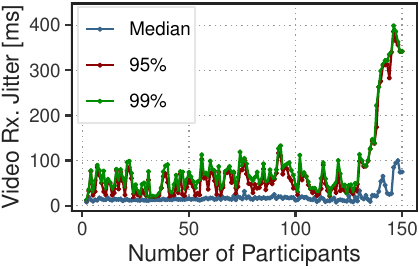}
        \caption{Video jitter while adding participants to the SFU.}
        \label{fig:rx-jitter}
    \end{minipage}
    \hspace{0.4cm}
    \begin{minipage}[t]{0.31\textwidth}
        \centering 
        \includegraphics[width=0.85\columnwidth]{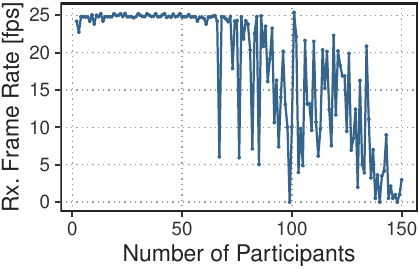}
        \caption{Video frame rate while adding participants to the SFU.}
        \label{fig:rx-fps}
    \end{minipage}
\end{figure*}


\subsection{Meeting Topologies and SFUs}%

\noindent
\textbf{The Need for SFUs.}
%
%
Modern video-conferencing applications use selective forwarding units (SFUs) to connect participants, making SFUs the de-facto standard for two main reasons:
First, while direct P2P connections are possible for meetings with more than two participants, they are impractical due to the need for each participant to encode and send media to every other participant, as depicted on the left in Figure~\ref{fig:call-arch}.
This results in significant computational overhead and requires substantial uplink bandwidth, which is often unavailable in residential and wireless Internet settings.
Second, using an intermediate (publicly routable) server solves challenges associated with firewalls and network address translators (NATs), making SFUs useful even for two-party calls.
For example, Google Meet always uses an SFU for two-party calls.


\noindent
\textbf{SFU Scaling Properties.}
%
%
While the use of SFUs solves the problem of constrained uplink bandwidth and reduces the required CPU resources at clients, SFUs do not entirely solve the scalability problem in multi-party video conferencing as sometimes suggested~\cite{webrtc-arch-basics-blog}.
It is important to note that the number of streams that are required in an SFU scenario still grows quadratically with the number of participants.
There are $N^2$ media streams for $N$ participants (per media type, i.e., video, audio, or screen share) in an SFU topology which even grows slightly faster compared to the $N(N-1)/2$ streams in a P2P topology.
The key difference is that all these streams are now sent to or received by the SFU, effectively moving the bottleneck from the clients to the SFU.
The second important observation is that the amount of work that an SFU needs to perform is determined by the number of media streams and not by the number of participants which therefore leads to quadratic scaling behavior at the SFU.
This is due to the fact that every incoming stream needs to be replicated for each of the downstream participants and then sent out.
Figure~\ref{fig:call-arch} shows an example of this with three participants where P2P connections result in a total of six streams while the SFU handles a total of nine streams.


\noindent
\textbf{SFU Load in Campus Trace.}
Of course, meeting participants do not share both audio and video at all times.
To get a better understanding of the actual number of streams per meeting in a real-world video-conferencing application, we analyzed Zoom usage data taken from the Zoom Account API~\cite{zoom-api} on our University campus.
More details on the data set can be found in Appendix~\ref{apx:zoom-api}.
Figure~\ref{fig:n-sfu-streams} shows the range (gray bars) and median number (blue dots) of media streams at the SFU for each meeting as a function of the maximum number of participants within each meeting on our campus.
A media stream was counted when it was active for at least 10\% of the meeting's duration.
The dashed line indicates the upper bound of streams possible if every participant shares both audio and video, which can be exceeded in practice when participants also share their screen, as seen in this figure. \sata{I wonder whether we should remove the effect of screenshare streams to keep things simple, and just mention at the end of this paragraph that we ignore screenshare in this analysis, but the same principles apply. I don't remember though how difficult it was to identify screenshare streams in our campus dataset.}
We can see that even for meetings with 10 participants, the SFU already handles up to 200 media streams.
Meetings with 25 participants, a typical classroom size, generate in excess of 700 media streams at the SFU in our data set and can theoretically produce up to 1250 streams.

\subsection{Consequences of Under-Provisioning}

\noindent
\textbf{SFU Performance Implications.}
%
%
In contrast to signaling and rate adaptation, which rely on packets and messages that are not part of the actual media streams, media distribution is latency-sensitive.
Naturally, the SFU must touch each of these media packets, and as such, any delay introduced by the SFU is added to the end-to-end delay a user experiences, directly impacting QoE.
Consequently, it is crucial that the forwarding delay and induced jitter are minimal.
However, software packet processing is subject to operating-system level delay artifacts stemming from scheduling, context switches, interrupts, etc., which add to the forwarding delay.
At hundreds of packets (typically between 800 and 1400 Bytes in size for video and around 200 Bytes in size for audio) per second per stream, SFU operations require copying significant amounts of data among socket buffers for receiving packets and before sending them out, resulting in high CPU load and frequent context switching.
These delays are hard to predict and can impair session quality to the point where the VCA becomes unusable.

\noindent
\textbf{QoE Degradation under high SFU Load.}
To confirm the suspected quality impacts of under-provisioning SFUs, we conducted an experiment using the Mediasoup open-source SFU~\cite{mediasoup} which we deployed on a server with a 40-core Intel Xeon Silver 4114 CPU and 96GB of RAM.
A second server with similar characteristics emulated clients using headless Chrome and was directly connected to the SFU via a 1Gbit/s Ethernet link.
We pinned the Mediasoup server to a single CPU and incrementally built up to 15 meetings with 10 participants each, adding one participant every ten seconds.
We measured the quality of the first meeting using the WebRTC statistics API~\cite{webrtc-stats} as we added more participants to the SFU.
During the experiment, the server reached 100\% CPU utilization at around 80 participants.
Figure~\ref{fig:rx-jitter} shows the receive jitter.
Tail jitter is high throughout the experiment before exceeding 100ms, causing significant mouth-to-ear delay or periodic freezes.
Figure~\ref{fig:rx-fps} shows that the video frame rate starts dropping at around 60 participants, with frequent drops beyond this point.
Overloading the SFU has direct and noticeable consequences for meeting quality, making it effectively unusable from around 100-120 participants.

\noindent
\textbf{Takeaways.}
%
%
In summary, while sharing scaling properties with other user-facing services such as diurnal usage patterns, SFUs have additional unique scaling properties due to the quadratically growing amount of media streams to be forwarded \sata{The sentence structure is a little convoluted -- could perhaps be simplified.}.
Dynamics and unpredictability within meetings, for example, due to participants joining or leaving or starting or stopping to share a particular media type (e.g., video, audio, or screen), further exacerbate this problem.
At the same time, under-provisioned SFUs can rapidly hit high utilization levels, which have a direct, noticeable, and sometimes prohibitive impact on the session quality.
Taken together, VCA operators either vastly over-provision their infrastructure to accommodate such dynamics or they jeopardize QoE.

%% file: sec/3-scallop.tex

\section{SFU as a Packet Processor}
\label{sec:sfus-as-packet-processors}

\noindent
Before presenting Scallop, we provide an overview of SFU operations and the key insights guiding our design.


\begin{figure}[t]
    \centering
    \includegraphics[width=\columnwidth]{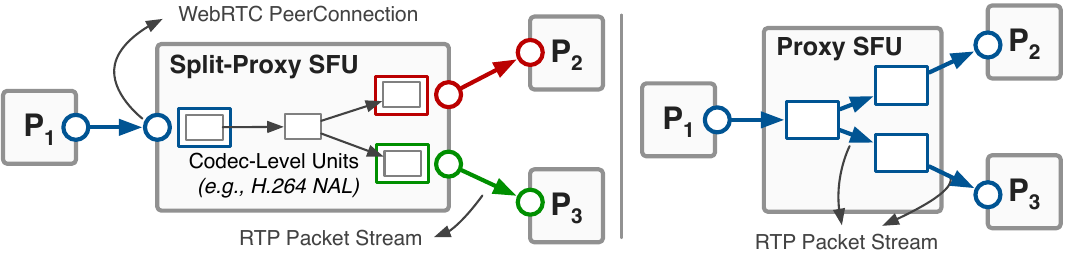}
    \caption{SFU Design Choices.}
    \label{fig:sfu-designs}
\end{figure}

\noindent
\textbf{SFU Design Choices.}
%
Today, SFUs are the de-facto standard in video conferencing.
In contrast to earlier VCA server designs, they operate on Real-Time Transport Protocol (RTP) packets containing media.
Even though WebRTC is the only widely adopted standard and open-source framework for video conferencing, it does not provide any guidance on how SFUs should be implemented or how they should handle RTP streams.
The simplest way to implement an SFU is to maintain separate P2P connections and distinct RTP streams between the SFU and each participant.
This approach is similar to a \emph{split-proxy} design, a term we will use going forward and is illustrated on the left side of Figure~\ref{fig:sfu-designs}. 
Existing WebRTC SFUs (e.g., MediaSoup~\cite{mediasoup}) operate in this way.

\noindent
\textbf{Media Forwarding in Zoom's SFUs.}
Previous work on Zoom, one of the market leaders in video conferencing, however, has shown that Zoom does not seem operate in this manner~\cite{michel:passive-measurement-zoom, how-zoom-works}.
We collected packet-level traces of real Zoom calls on our campus (see Appendix~\ref{apx:zoom-passive}) to confirm previous findings that Zoom's servers send out exact copies (except for rewritten IP addresses and port numbers) of the incoming RTP packets to all downstream participants.
While we do not know what architecture Zoom uses and how Zoom's SFUs are implemented, we observe that their media-forwarding operation is more akin to a traditional packet processor.
In this model, the required functionalities of replicating packets and rewriting header fields are supported even in older-generation SDN switches (e.g., those supporting OpenFlow).
This insight is the first foundation of our design for Scallop.

\noindent
\textbf{Rate Adaptation in Zoom's SFUs.}
Rate adaptation becomes necessary when the network conditions of a participant change, e.g., due to network congestion or if it is not necessary to forward high-resolution video to a participant due to device characteristics, such as a smaller screen on a cell phone.
Without rate adaptation at the SFU, media senders would all have to reduce their sending rate to relieve congestion, resulting in lower quality for all participants, even those unaffected by congestion.
Realizing this functionality (without transcoding media) requires using a scalable media stream.
Our analysis of Zoom's SFU revealed that it uses a combination of Simulcast and Scalable Video Coding (SVC) to achieve this.
In SVC, a video stream is encoded in multiple layers, each with a different resolution and bitrate.
The media stream is packetized so that a layer never crosses a packet boundary.
As a result, reducing the media resolution or frame rate can be achieved by dropping a specific subset of packets.
This insight is the second foundation of our design.
This concept is explained further and a real-world example of this functionality, as measured on Zoom traffic on our campus, is presented in Appendix~\ref{apx:zoom-simulcast}.

\noindent
\textbf{Takeaways.}
Taken together, the core work of SFUs can be implemented by replicating packets and sending copies out to all receiving participants.
Furthermore, if SVC (or the related technology, Simulcast) is used, adapting a media stream reduces to forwarding a clearly defined and correspondingly labeled subset of packets to a given participant.

\section{Introducing Scallop}%
\label{sec:scallop}

\noindent
Based on the insights from \S\ref{sec:sfus-as-packet-processors}, we introduce Scallop, a novel SFU design leveraging SDN principles and programmable networking hardware to improve the scalability and performance of video-conferencing infrastructure.
Scallop offloads all media replication, forwarding, and rate-adaptation tasks to high-speed hardware, yet several operations SFUs perform are not amenable for such an implementation.
Consequently, we require a split of functionality where we leave as many tasks as possible inside the data plane and only carefully leave operations in software when absolutely necessary.

\begin{figure}[t]
    \centering
    \includegraphics[width=0.95\columnwidth]{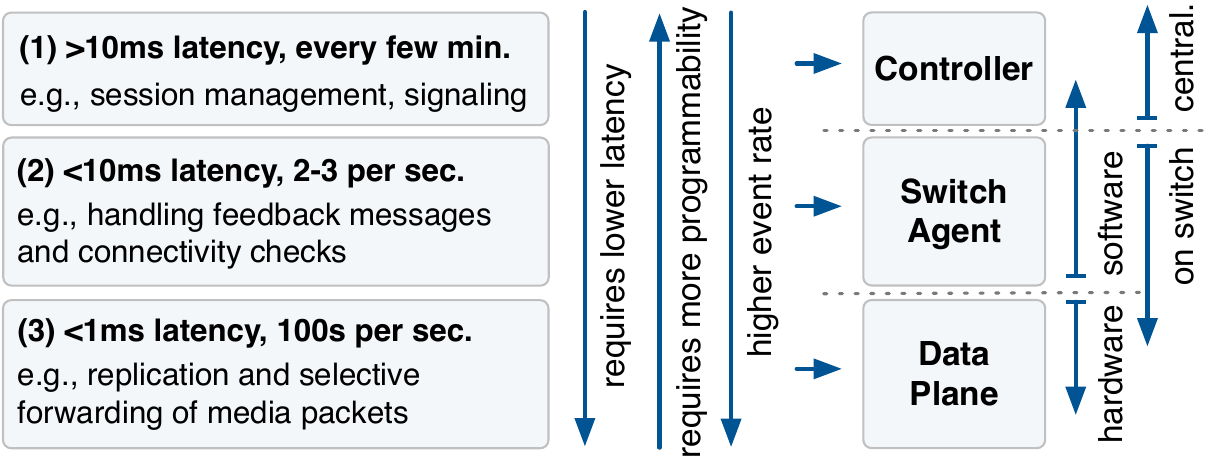}
    \caption{Latency and programmability requirements of key SFU responsibilities and resulting placement.}
    \label{fig:split}
\end{figure}

\noindent
\textbf{A Taxonomy of SFU Operations.}
Scallop's control/data-plane split and its corresponding system architecture are driven by the latency requirements, computational complexity, and frequency of the various tasks an SFU performs.
Along these axes, we classify SFU tasks into 3 categories as shown on the left side of Figure~\ref{fig:split}: (1) infrequent tasks that are not latency-sensitive, including session management and signaling; (2) latency-sensitive tasks (on the order of 10s of milliseconds), including deciding the correct target sending rate of a media stream to a participant based on periodic feedback signals, as well as handling periodic connectivity checks performed by the STUN protocol~\cite{stun-rfc}; (3) ultra-low-latency tasks (sub-millisecond), including the actual handling of media packets, i.e., forwarding and dropping them if necessary.
This taxonomy is illustrated on the left side of Figure~\ref{fig:split}.

\noindent
\textbf{Three-Tier SFU Design.}
The resulting split and architecture are illustrated on the right side of Figure~\ref{fig:split}.
This architecture is inspired by the design of SDN systems but goes a step further by introducing a third plane between a centralized controller and the switch data plane for the aforementioned latency-sensitive tasks.
In this architecture, the controller is only involved when (1) a new session is created, (2) a participant joins or leaves a meeting, or (3) a participant starts or stops sharing a particular media type (i.e., audio, video, or screen).
The controller benefits from a central view and, albeit not further described or implemented in this paper, is designed to control multiple SFUs in a \emph{cascading SFU} architecture (see Appendix~\ref{apx:additional-discussion}).
Finally, the data plane handles truly latency-sensitive tasks on the critical path for QoE.


\noindent
\textbf{Scallop's Switch Agent.}
All remaining tasks that are either (1) not amenable to implementation in the data plane due to their complexity, or (2) fall in the category of latency-sensitive tasks that not on the critical path for QoE, run in software directly on the switch (i.e., the switch's CPU and operating system).
This, for example, includes processing feedback signals that are subsequently used for rate adaptation.
We call this intermediate component the \emph{switch agent}.
Importantly, the switch agent is not involved in media forwarding, and none of the above-mentioned tasks require any media or feedback to be sent back from the switch agent; rather, the switch agent only receives copies of packets, analyzes them, and reconfigures the data plane if required.

%% file: sec/4-control.tex
\section{Control-Plane Prototype}%
\label{sec:control}

\begin{figure}[t]
    \centering
    \includegraphics[width=0.95\columnwidth]{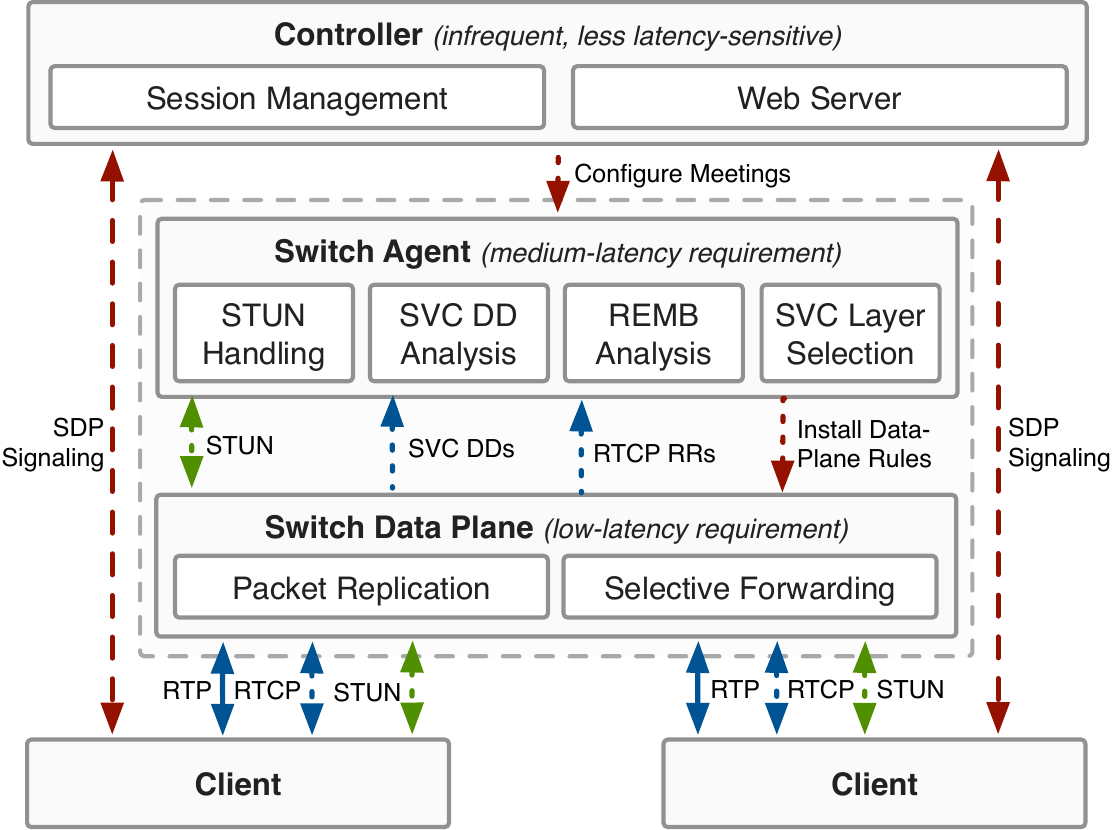}
    \caption{Scallop's 3-Tiered Architecture.}
    \label{fig:arch}
\end{figure}

%
\noindent
In order to realize Scallop's proxy-based SFU architecture, the control plane must establish and manage WebRTC sessions and handle the infrequent but important remaining tasks that cannot run in the data plane due to their complexity.
%
%
%
Scallop's control-plane architecture consists of two components: (1) a centralized controller and (2) a control program running on the switch CPU, which we call \emph{switch agent}.
Both applications are written in C++ and communicate with each other using remote procedure calls (RPC).
The switch agent can additionally exchange full packets directly with the data plane via the switch's CPU port.
This further subdivision between the centralized controller and the switch agent is depicted in Figure~\ref{fig:arch}.

\sata{Upon re-reading the sub-sections, I felt that we need a little more front-loading of the reasons each sub-section is important. A preamble to each sub-section that summarizes why the following information is relevant might be a good way to do that. Once we get into the details, because there are a lot of details, it's a little difficult to keep track of why.}


\subsection{Session and Connectivity Management}
\label{ssec:session-management}

%
\noindent\textbf{WebRTC Signaling.}~%
WebRTC uses the \emph{Session Description Protocol (SDP)} to negotiate media session parameters between participants, including codecs, their parameters, and IP addresses and ports~\cite{webrtc-signaling,rfc3264}.
This negotiation, known as \emph{signaling}, is initiated by participants whenever a new media stream is created.
Scallop's controller acts as the signaling server that exchanges SDP messages between participants and maintains state about participants and their media streams to correctly configure the data plane.


%
%
\noindent\textbf{Controlling Signaling to Create Proxy Topology.}~%
Each SDP message includes a list of \emph{connection candidates}, which convey the IP addresses and ports RTP media is being sent from or can be received at.
Using this information, we can \emph{insert} the SFU as an intermediate entity between participants while appearing to meeting participants as their sole peer.
\sys{} achieves this by intercepting SDP messages and modifying the connection candidates on the fly.

\noindent\textbf{STUN and Connectivity Management.}~%
After identifying potential participants via \emph{Interactive Connectivity Establishment (ICE)}~\cite{ice-rfc}, WebRTC uses periodic \emph{Session Traversal Utilities for NAT (STUN)}~\cite{stun-rfc} packets to continuously check reachability and also compute round-trip times between the participants. 
However, processing STUN packets in the data plane is too complex due to their header format. Thus, \sys{} handles STUN packets in the control plane.
This is fine since STUN packets are not classified as \emph{latency critical}---a connection is only deemed interrupted after  \textit{multiple consecutive connectivity checks} using STUN packets fail or time out.

\subsection{Bandwidth Estimation}
\label{ssec:bw-estimation}
\noindent
\sata{Even though we have said this before, I feel like we ought to remind readers here that rate adaptation happens in two parts: (1) the sender sends at the highest rate any of its receivers can receive at (5.2 and 5.3), (2) for receivers that can only support lower rates than the highest rate, the SFU scales down the stream by dropping packets (5.4).}
WebRTC uses Google Congestion Control (GCC)~\cite{carlucci:google-cc} to estimate the network capacity, which is used to adjust media stream bitrates accordingly.
In particular, we adopt GCC's receiver-driven mode, where the receiver estimates the bandwidth based on packet arrival time variation and \textit{periodically} sends these estimates to the sender using \emph{Receiver-Estimated Bandwidth (REMB)} messages.
Although REMB messages are still too complicated for the data plane to process, the frequency of them is proportional to the frequency of the link capacity changing, unlike the sender-driven mode, where the receiver sends a \emph{Transport-Wide-CC (TWCC)} message to the sender for every 10-20 media packets. \sata{Wonder whether we should start by saying there are two modes, and then explain why the receiver-driven mode suits us better. Otherwise, the modes seem to come up randomly.} We can also directly use the REMB value to decide the appropriate media bitrate to send at the SFU (i.e., performing the actual rate adaptation). 

\subsection{Preserving Feedback Semantics}
\label{ssec:preserving-feedback-semantics}
%
%
\noindent\textbf{Bandwidth Estimation in a Proxy Architecture.}~%
Realizing correct bandwidth estimation in our proxy-based Scallop architecture, however, is challenging.
This is because forwarding feedback among participants, rather than using individual control loops (as in a split proxy), mixes signals, incorporating uplink measurements from all senders instead of reflecting a specific downlink.
As a result, feedback converges on the lowest-bandwidth receiver, forcing the sender to lower the media bitrate unnecessarily for all participants.

\begin{figure*}[t]
    \centering
    \includegraphics[width=0.9\textwidth]{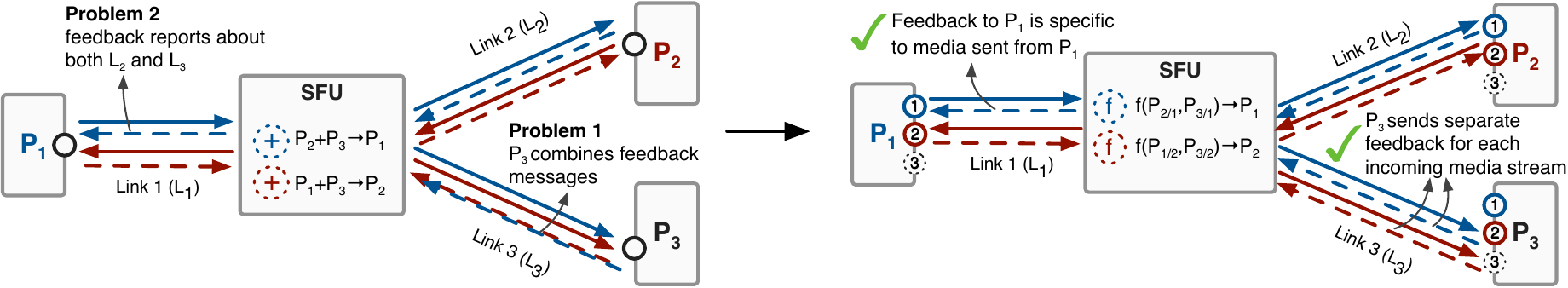}
    \caption{Splitting WebRTC connections per participant and forwarding feedback messages of the best-performing downlink only preserves feedback semantics and ensures effective rate adaptation.}
    \label{fig:rtcp-combined}
\end{figure*}


%
%
\noindent\textbf{Mixed Feedback Signals.}
To illustrate the problem, consider a 3-party meeting as depicted in Figure~\ref{fig:rtcp-combined}.
Participant 1 ($P_1$, blue) and 2 ($P_2$, red) share video while participant 3 ($P_3$) only receives.
Solid lines depict media streams and dashed lines depict feedback messages.
These feedback messages leverage the Real-Time Transport Control Protocol (RTCP).
We refer to a participant’s connection to the SFU as the uplink and the SFU’s connection to the participant as the downlink.
Done naively, two problems now arise.
First, RTCP combines feedback messages concerning multiple media streams into a single RTCP packet which is too complex to parse, analyze, and correctly forward in the data plane. 

\sata{Need to define RTCP since it comes up for the first time here.}
Second, they also mix feedback signals from different media streams in a way that feedback is not actionable anymore.
For example, the estimated bandwidth computed by $P_3$ now is based on feedback regarding both $P_1$'s and $P_2$'s uplinks in addition to $P_3$'s downlink.
This estimate is irrelevant for the SFU since it cannot do anything about uplink capacity.
More importantly, it also leads to the problem that when naively forwarding the combination of all messages to $P_1$, $P_1$'s media will now be encoded at a bitrate that caters also to $P_2$'s uplink performance.
Put differently, all send rates will converge to the lowest capacity of all uplinks in the meeting which defeats the purpose of having rate adaptation in the SFU in the first place.


\noindent\textbf{Solution Part 1: Split WebRTC Connections.}~%
Scallop solves these problems using two techniques, as depicted in Figure~\ref{fig:rtcp-combined}.
First, we split WebRTC UDP streams (which are the scope for bandwidth estimation) on a per-participant basis, such that feedback messages sent from each receiver are specific to one sender.
This frees Scallop from splitting combined RTCP feedback messages that report about multiple received streams to only forward the messages relevant to a particular sender to that sender.

\noindent\textbf{Solution Part 2: Selectively Forward Feedback.}
In our example, $P_3$ now has a dedicated stream to receive media from $P_1$ and a second stream to receive media from $P_2$.
Consequently, $P_3$'s REMB messages only include information about the path between $P_3$ and $P_1$ or $P_2$, respectively, and not anymore about the combination of all paths.
A filter function (denoted as $f$) at the SFU selects the best-performing downstream link per sender and configures the data plane to only forward these messages to the respective sender.
This filter function selects the best-performing downlink by computing an exponentially-weighted moving average (EWMA) over each receiver's bandwidth estimates.
It then periodically selects the maximum out of the EWMAs.
If the maximum is different from the previous maximum and is associated with a different sender, Scallop reconfigures the data plane to only forward REMB messages from this receiver to the respective stream's sender.
The rationale behind this is that the sender should transmit at the highest rate allowed by its uplink and the best downlink.
Since all packets traverse the sender's uplink, it is inherently accounted for in the feedback.
The SFU then handles adaptation for lower-bandwidth downlinks, as detailed in the next section.

\subsection{SVC Analysis and Layer Selection}
\label{ssec:svc-analysis-and-layer-selection}

\begin{figure}[t]
    \centering
    \includegraphics[width=0.9\columnwidth]{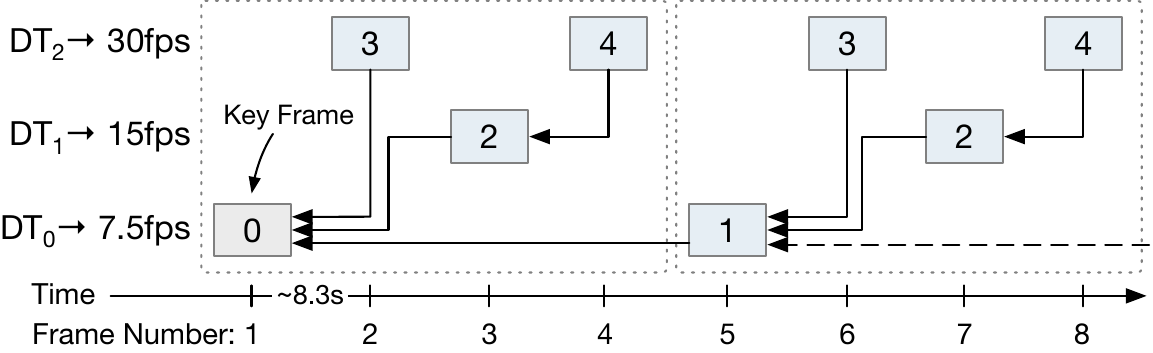}
    \caption{Frame dependencies in AV1 L$_1$T$_3$ SVC.}
    \label{fig:l1t3}
\end{figure}

\noindent\textbf{Scalable Video Coding.}~%
\sys{} leverages \emph{Scalable Video Coding (SVC)}~\cite{schwarz:svc-overview} with the AV1 codec's L$_1$T$_3$ profile~\cite{av1,av1-rtp} to adapt media streams to network capacity. 
This allows the SFU to choose among three temporal layers (i.e., frame rates) for video streams.
While the data plane handles the actual forwarding, the control plane decides which quality level to send to each participant.

\noindent\textbf{AV1 Dependency Descriptors.}~%
The SVC structure can change with each key frame (sent when a stream starts or the resolution is changed), which requires the SFU to adapt the data plane's forwarding rules accordingly.
Figure~\ref{fig:l1t3} shows the dependencies in an L$_1$T$_3$ SVC stream.
Key frames contain a dependency descriptor that carries this information.
In AV1, each RTP packet contains an RTP extension header indicating its layer through a unique \emph{template id}.
Template ids 0 and 1 represent the base layer (7.5 fps), id 2 the first enhancement layer (15 fps), and ids 3 and 4 the second enhancement layer (30 fps).
Dropping frame ids 3 and 4 would reduce the frame rate from 30 fps to 15 fps.
The data plane parses the AV1 extension header to decide whether the packet should be dropped or not.
Appendix~\ref{apx:rtp-extension-parsing} describes the parsing process and its associated challenges.

\noindent\textbf{Selecting a Quality Layer at \sys{}'s Control Plane.}~%
Parsing the AV1 header \sata{(contained inside the AV1 extension header)} is beyond the data plane's capabilities. Thus, \sys{} sends key frames to the control plane for analysis. \sata{The REMB point comes up a little abruptly.}
Whenever a REMB message is received from a participant, the switch agent invokes a function that can be defined by adopters of Scallop.
This function is declared as follows: \\
\texttt{\small selectDecodeTarget(currDT, estHist, newEst) $\rightarrow$ newDT}.\\
The function takes as input the current decode target (currDT), a history of past estimates (estHist), and the new estimate (newEst) from the REMB message; it returns the new decode target.
If the returned decode target is different from the previous one, the switch agent then reconfigures the data plane.
Importantly, while we implemented a simple heuristic that switches between quality levels based on fixed capacity-estimate thresholds, using this model, arbitrary rate-adaptation algorithms can be implemented.


\subsection{Handling other RTCP Messages}
\label{ssec:handling-of-other-rtcp-messages}

\begin{figure}[t]
    \centering
    \includegraphics[width=0.85\columnwidth]{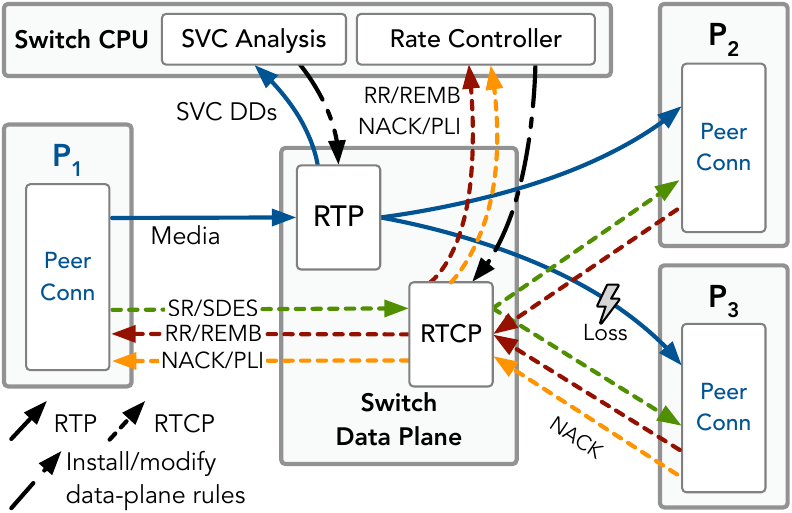}
    \caption{Flow of all types of media and control packets in a Scallop 3-party conference.}
    \label{fig:rtcp-fwd}
\end{figure}


\noindent
Apart from REMB, there are additional feedback messages delivered through RTCP. 
From the receiver side, negative acknowledgments (NACK) request the retransmission of a specific media packet, and picture-loss indication messages (PLI) notify the sender to send an intra-coded video frame.
As shown in Figure~\ref{fig:rtcp-fwd}, NACK and PLI messages are sent from a receiver experiencing loss and are forwarded to the respective sender.
While PLIs, NACKs, and REMB messages are forwarded to their intended destinations through the data plane without delay, the data plane also creates copies of them on-the-fly, sending them to the switch CPU for further analysis (e.g., for running the filter function described earlier).
On the sender side, sender reports (SR) periodically time-synchronize streams originating at a sender, and also report various statistics to all receivers.
SR messages are depicted as green dashed arrows in Figure~\ref{fig:rtcp-fwd}.

%% file: sec/5-data-plane.tex
\section{\sys{} Data Plane}
\label{sec:data-plane}

\noindent
In this section, we describe how \sys's data plane performs its two main tasks---scalable media replication (\S\ref{ssec:data-plane-media-replication}) and transparent rate adaptation (\S\ref{ssec:data-plane-transparent-rate-adaptation})---followed by our hardware prototype implemented with the Intel Tofino2 switch (\S\ref{ssec:data-plane-hardware-prototype}).

\subsection{Scalable Media Replication}
\label{ssec:data-plane-media-replication}

\begin{figure*}[t]
    \centering
    \includegraphics[width=0.9\textwidth]{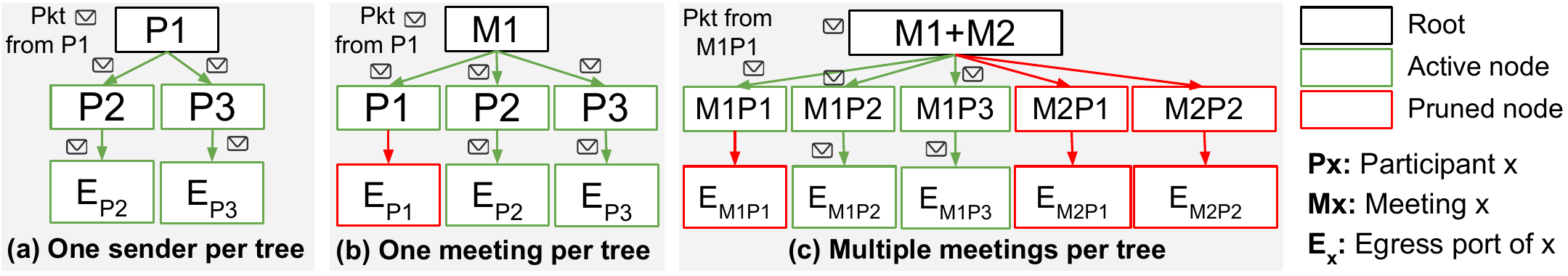}
    \caption{Constructing efficient replication trees by aggregating meetings and leveraging dynamic pruning.}
    \label{fig:replication-trees}
\end{figure*}

\boldpara{Background and challenges.}
\sys{} needs to replicate media packets at scale to serve thousands of concurrent meetings and participants. Modern programmable networking hardware supports \emph{highly-configurable packet replication engines}, presenting us with an opportunity to achieve this. In one common design of a replication engine, an incoming packet arrives at a \emph{root} node, which replicates it to multiple level-1 (L1) nodes; each L1 node replicates again to level-2 (L2) nodes, and each L2 node directs the final replicas to corresponding egress ports (\S\ref{ssec:data-plane-hardware-prototype}). These engines also enable dynamic \emph{pruning} of nodes at both L1 and L2 based on rules set by the control plane and matched in the data plane, allowing unwanted replicas to be \emph{suppressed}.

However, while highly configurable, replication engines are primarily designed for traditional multicast, where receivers join a multicast group and routers replicate and forward \emph{identical} packets to them in a one-to-many model. In contrast, VCA SFUs need to support per-receiver rate adaptation and operate in unicast-only networks, introducing additional complexity. No prior work has systematically mapped SFU-style replication to these hardware engines. Thus, \sys{} explores the design space from the ground up. Our key challenges include deciding how to map VCA entities (meetings, senders, receivers) onto the replication hierarchy (root, L1 node, L2 node) and managing resource constraints, such as a fixed maximum number of replication trees ($\mathcal{T}$) and constrained flexibility to prune undesired multicast branches.
A naive \emph{one tree per media stream} approach (Figure~\ref{fig:replication-trees}(a)) quickly depletes $\mathcal{T}$ replication trees and fails to utilize the switch’s high bandwidth.
To overcome this, \sys{} carefully constructs and aggregates multicast trees to fit as many meetings and participants as possible in the replication engine, dynamically prunes branches to replicate only the packets actually needed downstream, and preserves rate adaptation by selectively replicating SVC layers.

\boldpara{Efficient replication for non-rate-adapted (NRA) meetings.}
For meetings that do not require rate adaptation (i.e., all participants can send and receive media at the highest quality), \sys{} first aggregates all participants of a single meeting into a single replication tree, as shown in Figure~\ref{fig:replication-trees}(b). Instead of allocating a separate tree for each sender (as in the naive design), the root now represents the entire meeting, with L1 nodes representing each participant and L2 nodes mapping participants to their egress ports. This design supports $\mathcal{T}$ concurrent meetings and reduces the per-meeting replication overhead from $\mathcal{N}(\mathcal{N}-1)$ L1 nodes (in the naive approach) to $\mathcal{N}$ L1 nodes, where $\mathcal{N}$ is the number of participants in a meeting. However, pruning at L2 is required to prevent senders from receiving their own packets (we describe the exact pruning process in \S\ref{ssec:data-plane-hardware-prototype}).
To increase efficiency further (from one meeting per replication tree), \sys{} maps multiple meetings (say, $m$ meetings) into a single replication tree (Figure~\ref{fig:replication-trees}(c)), supporting $m\mathcal{T}$ concurrent meetings, where $m$---the number of meetings per tree---is a constant determined by the capabilities of the hardware platform. However, in addition to pruning at L2, this approach requires pruning at L1 to ensure that packets of a meeting are not received by participants in a different meeting. For example, in Figure~\ref{fig:replication-trees}, when a packet arrives for meeting M1, the replicas to M2's participants are suppressed, and vice-versa. Using this NRA design, our prototype can support up to 128K concurrent meetings (\S\ref{subsec:data-eval}).

\boldpara{Efficient replication for receiver-specific rate-adapted (RA-R) meetings.}
\sys{} needs to selectively replicate SVC layers to receivers to support rate adaptation (\S\ref{ssec:svc-analysis-and-layer-selection}). This requirement complicates replication-tree design since there are constraints on pruning branches in hardware. The control plane can only update trees infrequently to remain scalable, while the data plane is limited by the amount of hardware-internal metadata it can carry for pruning decisions. This forces a choice between pruning either a subset of participants or a subset of SVC layers, but not both simultaneously. In a common scenario where each receiver is assigned the same media quality (e.g., 15 fps video) across all senders (i.e., receiver-specific rate adaptation), \sys{} addresses these limitations by creating one replication tree per media quality. For $L_1T_3$, \sys{} creates three trees (for low, mid, and high qualities) per $m$ meetings. With $q$ qualities, the system can thus support $m\mathcal{T}/q$ meetings, requiring up to $t\mathcal{N}$ L1 nodes per tree. Using this RA-R replication-tree design, our prototype can support up to 42.7K concurrent meetings (\S\ref{subsec:data-eval}).

\boldpara{Efficient replication for sender-receiver-specific rate-adapted (RA-SR) meetings.}
For rate adaptation scenarios where a receiver's media quality may differ across different senders, \sys{} cannot do better than aggregating two senders (and their corresponding receivers) per media quality into a single replication tree. Under this RA-SR design, \sys{} supports up to $2\mathcal{T}/q\mathcal{N}$ concurrent meetings while requiring up to $2\mathcal{T}$ L1 nodes. Using this design, with 10 participants per meeting (all sending video and audio), our prototype can support up to 4.3K concurrent meetings, compared to 192 supported by a 32-core server.

\boldpara{Handling two-party meetings efficiently.}
While optimizing resource utilization for \emph{multi-party} meetings is key to \sys's efficiency, \emph{two-party} meetings are quite common ($60\%$ meetings in our campus dataset). Since two-party meetings do not require replication, we further optimize our design by not constructing replication trees for such meetings and treating the sender's media stream as unicast. With this optimization, when all meetings are two-party, our prototype supports up to 533K concurrent meetings, compared to 4.8K supported by a 32-core server.

\boldpara{Dynamic migration across replication tree designs.}
Since rate adaptation is not always required, and even when required it may not be sender-receiver-specific, we design the switch agent to dynamically migrate meetings between two-party, NRA, RA-R, and RA-SR designs as needed. This maximizes the utilization of the replication engine, ensuring that the bottleneck remains the switch’s bandwidth rather than replication capacity. To avoid meeting disruptions during migration, we: (1) create a new replication tree for the migrated meeting, (2) update data plane rules to redirect packets to the new tree, and (3) deallocate the old replication tree.

\boldpara{Addressing replicated packets.}
Unlike traditional multicast, where all replicas share the same source and multicast destination, an SFU must send each packet to a specific unicast receiver in keeping with the semantics of a WebRTC-compliant split-proxy design. To achieve this, \sys{} rewrites packet headers in the egress pipeline, setting the SFU's address as the source and the receiver’s unicast address as the destination, before forwarding it to the correct port.

\subsection{Transparent Rate Adaptation}
\label{ssec:data-plane-transparent-rate-adaptation}

\begin{figure}[t]
    \centering
    \includegraphics[width=0.75\columnwidth]{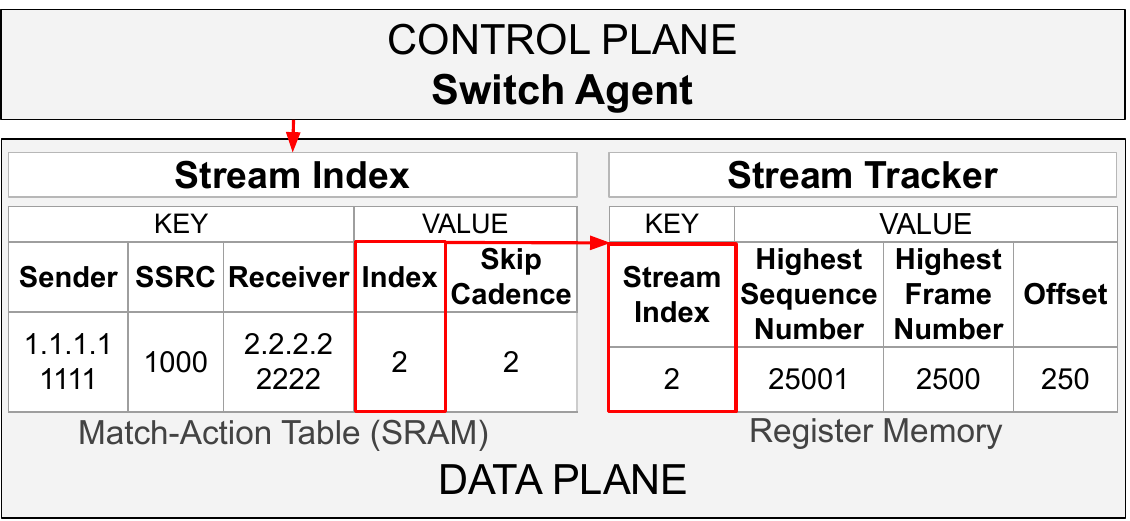}
    \caption{Rewriting sequence nos. for rate adaptation.}
    \label{fig:seqnum-rewrite}
\end{figure}

\boldpara{Background and challenges.}
Under WebRTC’s P2P model in a split-proxy architecture, each receiver expects a continuous, unmodified media stream from the SFU. However, in our true-proxy architecture, \sys{} performs rate adaptation by suppressing certain packets (in fact, with SVC, it suppresses entire media qualities containing multiple packets) to match network conditions. This suppression creates gaps in RTP sequence numbers, which receivers interpret as packet losses, triggering unnecessary retransmissions. To preserve the illusion of a direct stream by masking \emph{intentional} gaps, \sys{} needs to rewrite sequence numbers post-replication. However, rewriting sequence numbers perfectly---when suppression coincides with network-induced loss and reordering---is impossible even in software. \sys{} has to rely on a \emph{heuristic} instead, which can occasionally make mistakes, triggering unnecessary retransmissions. However, our experiments reveal that some mistakes are costlier than others: if we erroneously \emph{leave a sequence gap}, the receiver only triggers retransmissions for the packets it perceives as missing, but if we \emph{duplicate} sequence numbers, the decoder's state breaks and the video freezes indefinitely. Therefore, \sys{} \emph{always} avoids duplicating sequence numbers. Finally, \sys{} rewrites sequence numbers in a hardware-amenable manner, i.e., in a streaming fashion with a resource footprint low enough to fit in the programmable switch pipeline.
We implemented multiple such heuristics, each with different trade-offs between erroneous retransmission overhead and memory footprint, two of which we discuss below.

\boldpara{(1) Sequence Rewrite--Low Memory (S-LM): }
\sys{} keeps a \emph{Stream Tracker} hash table in the data plane, storing the highest observed sequence number, the highest observed frame number, and the sequence number offset for each active rate-adapted stream (Figure~\ref{fig:seqnum-rewrite}). The sequence number offset is the difference between the original sequence number and the rewritten sequence number. A stream is identified by its sender’s address, receiver’s address, and synchronization source (SSRC). The control plane provides the \emph{frame-skip cadence} (e.g., every third frame is suppressed) so the data plane can suppress the corresponding packets. The control plane also provides a unique, collision-free hash-based index for each new stream via match-action rules in a \emph{Stream Index} table: this helps use the Stream Tracker table's memory maximally. Upon receiving the first packet of a stream, \sys{} records its sequence and frame numbers in the Stream Tracker table.
For each subsequent packet, \sys{} does the following:
\begin{enumerate}[leftmargin=*]
    \item Consecutive sequence number: Rewrite using offset.
    \item Skipped sequence numbers: Check if the gap matches the skip cadence. If so, increment the offset accordingly.
    \item Older (reordered) packet: If it is exactly one less than the last observed sequence number, rewrite it using the current offset. Otherwise, drop it to avoid possibility of sequence number duplication.
\end{enumerate}
This approach has a relatively low memory footprint and is well-suited for environments with low loss and reordering.

\boldpara{(2) Sequence Rewrite--Low Retransmission (S-LR):}
For environments with higher loss and reordering, \sys{} stores additional state: the first and highest sequence numbers of the latest observed frame, a boolean indicating whether the last frame ended, and the highest suppressed frame number. With such extra information, \sys{} is able to deal with a wider range of cases involving lost or out-of-order packets, reducing retransmission overhead. This variant lowers retransmission overhead by consuming more memory than the S-LM variant.
We evaluate the impact of the two heuristics on scalability in Section~\ref{sec:eval}.

\begin{figure}[t]
    \centering
    \includegraphics[width=0.8\columnwidth]{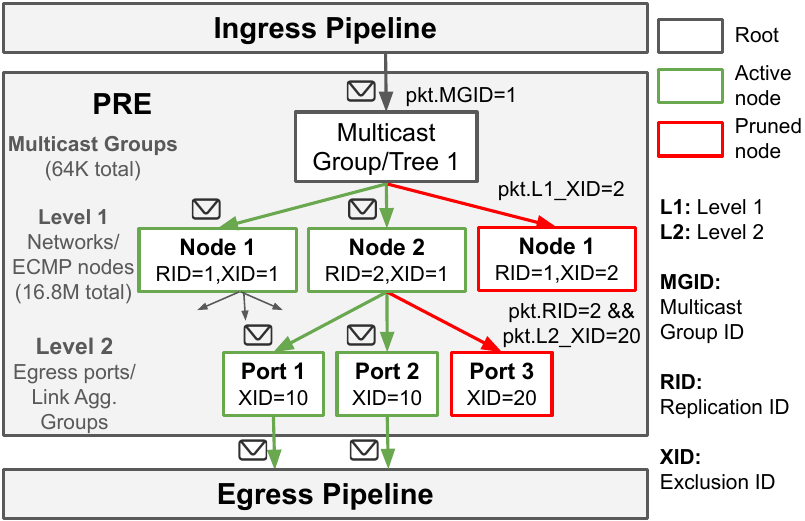}
    \caption{Tofino's Packet Replication Engine (PRE).}
    \label{fig:tofino-pre}
\end{figure}

\subsection{Hardware Prototype}
\label{ssec:data-plane-hardware-prototype}

\boldpara{Configuring the Tofino PRE.}
The PRE is a specialized hardware block in Tofino designed for efficient multicast through a hierarchical, three-level structure (called a \emph{multicast group} or a \emph{multicast tree}) managed by the control plane (Figure~\ref{fig:tofino-pre}). The PRE can support up to 64K multicast trees and a total of 16.8M ($2^{24}$) L1 nodes.
Tofino's PRE also supports dynamic pruning of branches (red arrows) using L1- and L2-exclusion IDs (XIDs).
The switch control plane configures the PRE at runtime by first creating L1 nodes, each identified by a \emph{node ID} (unique across the PRE), \emph{replication ID} or RID (unique across one multicast tree), and \emph{egress port} (unique across one RID in a tree). The control plane then defines multicast trees, each with a unique MGID and an associated set of L1 nodes. Each L1 node is assigned an L1 XID (the same XID can be assigned to multiple nodes) and a flag indicating whether pruning (via the XID) is enabled. Additionally, the control plane maps each L2 XID to a specific subset of the switch's egress ports. The data plane invokes the PRE by assigning the packet's MGID, L1 XID, and RID--L2 XID pair in the ingress pipeline. We aggregate two meetings in the same tree (i.e., $m$=2) by setting L1-XID=1 for the participants of the first meeting and L1-XID=2 for those of the second. When a packet from the first meeting arrives, we set its L1-XID to 2 in the data plane to exclude the second meeting's participants, and vice-versa. In addition, we suppress copies to the sender by setting the packet's RID to that of the sender and its L2-XID to its egress port. We identify replicated packets from the same sender uniquely in the egress pipeline by their RIDs (up to 65,536 RIDs are available for use per tree).

\boldpara{Per-stream state.} The egress pipeline stores six hash tables (\S\ref{ssec:data-plane-transparent-rate-adaptation}), always accessed in order, to enable sequence number rewriting on the Tofino. Each active rate-adapted stream maintains state across all six tables. Since the indices are managed by the control plane ensuring zero hash collisions and immediate cleanup when a stream ends, all cells in these tables can be used, allowing up to 65,536 concurrent streams.

%% file: sec/6-eval.tex
\section{Evaluation}%
\label{sec:eval}

\begin{figure*}[t]
    \centering
    \begin{subfigure}[t]{0.32\textwidth}
        \centering
        \includegraphics[width=0.8\textwidth]{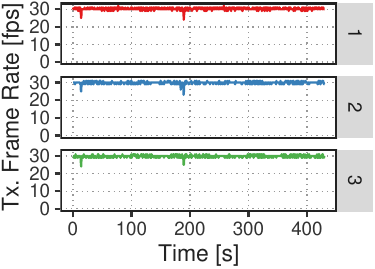}
        \caption{Send Frame Rate}
    \end{subfigure}%
    ~ 
    \begin{subfigure}[t]{0.32\textwidth}
        \centering
        \includegraphics[width=0.8\textwidth]{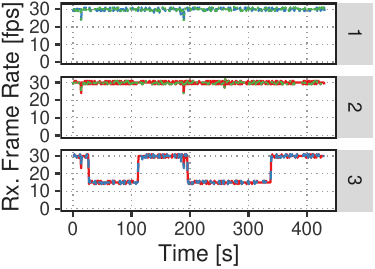}
        \caption{Receive Frame Rate}
    \end{subfigure}
    ~ 
    \begin{subfigure}[t]{0.32\textwidth}
        \centering
        \includegraphics[width=0.8\textwidth]{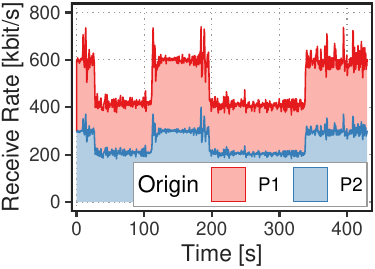}
        \caption{Receive Bit Rate at Part. 3}
    \end{subfigure}

    \caption{Example of \sys-Based Rate Adaptation: Participant 3's Receive Bitrate is Reduced Twice.}
        \label{fig:fidelity}
\end{figure*}


\noindent



\noindent
\textbf{Experimental Setup.}~%
We deploy the data plane of \sys{} on an 12.8 Tbit/s Intel Tofino2 hardware switch in our campus testbed.
The switch agent runs on the CPU of this switch, an 8-core Intel Pentium with 8GB of RAM.
The controller is deployed on a 40-core Intel Xeon server with 96 GB memory.
We use another server with the same configuration for both Scallop and MediaSoup (where applicable) clients where each participant runs in a Chrome browser tab.



\subsection{Control Plane}
\label{subsec:control-eval}

\noindent
We first analyze the amount of packets and bytes that our controller needs to process compared with the amount of packets that stay entirely in the data plane.
This ratio demonstrates the feasibility of Scallop's control/data-plane split.
We collect a packet-level trace of a real three-party meeting using \sys{} where participants send a 720p AV1 video stream using SVC with two temporal layers and an audio stream.
We then analyze the trace to determine the number of packets and bytes that can be processed in the data plane and the control plane, respectively.
The experiment ran for ten minutes during which we collected a total of 180,718 packets.

\begin{table}[t]
    \centering
    \scriptsize
    \begin{tabular}{l|rrr|rr} 
        \textbf{Proto./Type} & \textbf{Packets} & \textbf{Pct.} & \textbf{Per sec.} & \textbf{KBytes} & \textbf{Pct.} \\
        \hline
        RTP & 170,870 & 94.5 & 284.30 & 166,762 & 99.47 \\
         - Audio & 29,746 & 16.46 & 49.49 & 3,826 & 2.28 \\
         - Video & 141,124 & 78.09 & 234.81 & 162,935 & 97.19 \\
         \hspace{0.1cm} - AV1 DS* & 5 & $\lll$ & 0.008 & 6 & $\lll$ \\

        RTCP & 9,153 & 5.06 & 15.22 & 801 & 0.48 \\
         - SR/SDES & 3,456 & 1.91 & 5.75 & 304 & 0.18 \\
         - RR* & 240 & 0.39 & 0.13 & 15 & 0.01 \\
         - RR/REMB* & 5,457 & 3.02 & 9.07 & 482 & 0.29 \\
        
         STUN*& 695 & 0.38 & 1.15 & 89 & 0.05 \\
        \hline
        \textbf{Ctrl. Plane} & 6397 & \textbf{3.54} & 10.64 & 593 & \textbf{0.35} \\
        \textbf{Data Plane} & 174,326 & \textbf{96.46} & 290.06 & 167,066 & \textbf{99.65} \\
        \hline
        Total & 180,718 & 100 & 300.69 & 167,653 & 100 \\
    \end{tabular}
    \vspace{0.2cm}
    \caption{Packets per participant sent to SFU (10 min.)}
    \label{tab:pkt-types}
        \vspace{-0.3cm}
\end{table}

The results of this analysis are summarized in Table~\ref{tab:pkt-types}.
We found that 94.5\% of these packets were RTP packets which can be handled in the data plane with the exception of five RTP packets containing an extended AV1 dependency descriptor.
RTCP accounted for 5.06\% of all packets and 0.48\% of all bytes.
\sata{out of which,} Our switch agent uses RTCP receiver reports and REMB messages to control the rate-adaptation logic.
These packets accounted for 3.41\% of overall packets.
Finally, STUN packets accounted for 0.38\% of all packets and also need to be processed in software.
In summary, 96.46\% of all packets and 99.65\% of all bytes can be processed in the data plane, showing that this workload is naturally suited for a control/data-plane split design.
Almost more importantly, the remaining packets, with the exception of a few STUN packets at the beginning of each session for connection establishment, are not blocking.

\subsection{Data Plane}
\label{subsec:data-eval}

\noindent
\textbf{Tofino resource utilization.}
We implement \sys's data plane using $\sim$2000 lines of P4$_{16}$ code. We report the performance of our prototype in Table~\ref{tab:tofino-usage} in appendix F (the = symbol in the last column implies that the resource usage is same under any setting). We categorize the resource types available on the Tofino by how they scale with the number of participants---they either remain fixed, scale linearly, or quadratically. 
The number of stages consumed by our program falls squarely within the bounds of the maximum available on the Tofino.
For the rest of the components with fixed scaling behavior, we report the average utilization across all stages. We observe that the resource usage is low enough (always <22\%) such that other network applications can be comfortably deployed in conjunction with \sys.

\begin{figure}[t]
    \centering
    \begin{minipage}[t]{0.22\textwidth}
        \centering
        \includegraphics[width=\columnwidth]{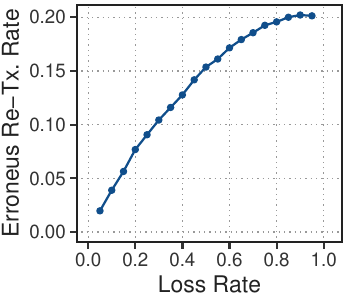}
        \caption{Retransmission Overhead with S-LR.}
        \label{fig:seq-rewrite-rtx-rate}
    \end{minipage}
    \hspace{0.2cm}
    \begin{minipage}[t]{0.22\textwidth}
        \centering %
        \includegraphics[width=\columnwidth]{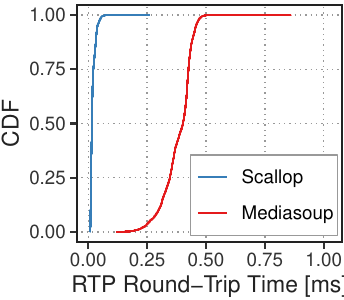}
        \caption{Scallop Forwarding Latency}
        \label{fig:rtp-rtt}
    \end{minipage}
\end{figure}

\noindent
\textbf{Overhead due to sequence number rewriting.}
In Figure~\ref{fig:seq-rewrite-rtx-rate}, we report the overhead of our sequence number rewriting heuristic S-LR, which is designed for a high-loss environment (\S\ref{ssec:data-plane-transparent-rate-adaptation}). This overhead is in terms of the fraction of extra retransmissions triggered by the receiver due to a mismatch between the rewritten sequence number and the ideal rewritten sequence number an oracle may have generated. We observe that the overhead is below 5\% for up to 10\% loss rate, and around 7.5\% for a 20\% loss rate. Even under really high loss rates (where the meeting itself would start to become unusable), the overhead does not exceed 20\%.


\begin{figure*}[t]
    \centering
    \begin{minipage}[t]{0.32\textwidth}
        \centering 
        \includegraphics[width=0.9\columnwidth]{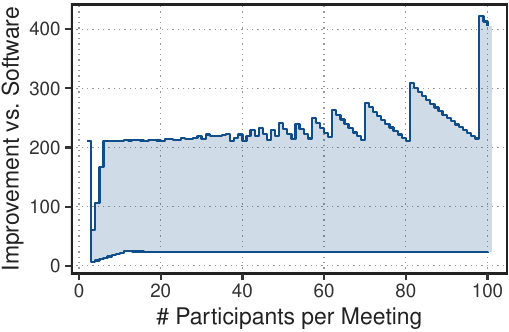}
        \caption{Scallop Scalability Gain over Software}
        \label{fig:sw-gain}
    \end{minipage}
    \hspace{0.2cm}
    \begin{minipage}[t]{0.32\textwidth}
        \centering 
        \includegraphics[width=0.9\columnwidth]{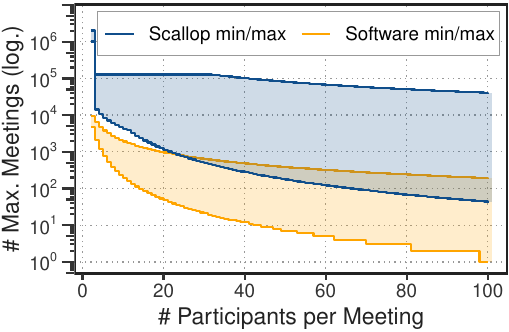}
        \caption{Best-case and worst-case Performance.}
        \label{fig:best-worst}
    \end{minipage}
    \hspace{0.2cm}
    \begin{minipage}[t]{0.32\textwidth}
        \centering 
        \includegraphics[width=0.9\columnwidth]{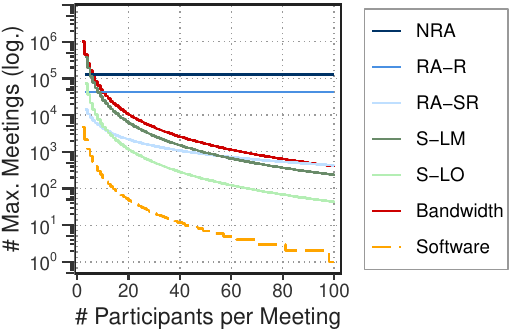}
        \caption{Performance of Replication-Tree Construction Designs.}
        \label{fig:repl-tree-perf}
    \end{minipage}
\end{figure*}

\subsection{Latency and Impact on Session Quality}

\noindent
We demonstrate that leveraging a hardware-based data plane significantly reduces SFU-induced delay. This is shown by comparing per-packet RTTs of RTP media packets in a two-party call.
Two participants are connected either through \sys{}'s Tofino or the Mediasoup's SFU server. 
Figure~\ref{fig:rtp-rtt} shows that \sys{} achieves 26.8$\times$ lower median latency while cutting 99\%ile latency by 8.5$\times$.

Additionally, to show that \sys{} is faithful to the core SFU functionality, we validate that \sys{}'s SVC-based rate adaptation effectively reduces bitrate without causing freezes. To confirm this, we start a three-party call, where all participants send and receive video. We make one participant emulate a poor downstream connection, requiring adaptive bitrate reduction of the streams it receives from other participants. In Figure~\ref{fig:fidelity}, we plot the receive frame rates measured by the media decoders at this participant, as an accurate representation of actual playback quality. The results confirm that \sys{} successfully reduces the frame rate from 30 to 15 fps for the constrained participant while maintaining a decodable media stream without incurring otherwise lower playback quality (e.g., through reduced resolution or freezes).



\subsection{Scalability}
\label{subsec:scale-eval}

\noindent
\textbf{Bytes processed by the switch agent.}
As we explained in Section~\ref{subsec:control-eval}, only a tiny fraction of packets and bytes are processed in software when \sys{} is used, as opposed to a software-based SFU. Figure~\ref{fig:switch-agent-bw} in appendix~\ref{apx:zoom-passive} also makes this point visually.

\noindent
\sata{I think we need to explain the simulation we ran to get these numbers. It's ambiguous otherwise how we got these numbers.}
\textbf{Scalability improvement over software.}
In Figure~\ref{fig:sw-gain}, we show the improvement in scale gained from using Scallop over a 32-core server, as the ratio of the number of meetings supported by Scallop as opposed to the server. The blue range shows that Scallop can support anywhere between 7-210$\times$ more meetings than software, depending on the specific meeting configurations (e.g., number of participants in a meeting, number of senders among those participants, etc.) and the variant of Scallop used (e.g., RA-SR vs. NRA, S-LR vs. S-LM). We obtain significant gains in case of two-party meetings due to our optimization discussed in \S\ref{ssec:data-plane-media-replication}. Thereafter, the improvement grows linearly since Scallop scales linearly while software scales quadratically. Finally, the lower bound of the blue region indicates the memory bottleneck encountered when S-LR is used, while the upper bound shows that this bottleneck is less impactful with S-LM. In Figure~\ref{fig:best-worst}, we show the range of meeting counts supported by Scallop as opposed to a 32-core server. The upper bound of each reflects the number of supported meetings when only one participant out of N (where N is the value on the x-axis) send media (video+audio); the lower bound reflects the same when all N participants send media simultaneously. We observe that, for given number of senders, Scallop always supports many more meetings than software, irrespective of the total number of participants in the system, the number of participants in a meeting, or the number of senders.

\noindent
\textbf{Software vs. hardware bottlenecks.}~%
Scallop presents multiple ways to construct efficient multicast trees in our hardware packet-replication engine (NRA vs. RA-R vs. RA-SR) as well as multiple sequence number rewriting heuristics (S-LM vs. S-LR).
The different designs have varying trade-offs that depend on the number of participants per meeting as well as how many of them send.
Figure~\ref{fig:repl-tree-perf} shows Scallop's performance if the respective hardware limitation was the bottleneck with all participants in a meeting always sending media.
The overall system performance becomes the minimum of all these lines.
In orange we plot the performance of a 32-core server for comparison.

%% file: sec/7-discussion.tex
\section{Discussion}
\label{sec:discussion}
\boldpara{Encryption and message authentication.}~%
In WebRTC, RTP headers are HMAC-protected while payloads are also AES-encrypted via SRTP-DTLS~\cite{srtp-rfc}. 
In \sys, neither mechanism is currently implemented.
The main challenge is key distribution: WebRTC’s split-proxy design uses a separate key for each P2P connection, exchanged between SFU and client.
This is incompatible with the proxy redesign where a single sender’s packets are replicated to many receivers, necessitating one-to-many key distribution, for example via centrally distributed keys (as previously done in WebRTC via SDES~\cite{webrtc-security} and done today in Zoom~\cite{how-zoom-works}).
Since the SFU does not need to touch payloads, it would then operate on encrypted packets.
Unlike in SRTP-DTLS, doing so would also be compatible with end-to-end encryption (E2EE) schemes.
Prior work demonstrates that ciphers (e.g., AES) and cryptographic hashes (e.g., SipHash) can be computed on programmable data-plane devices~\cite{yoo2021secure,oliveira2021dh,chen2020implementing,schulz2023cryptographic,scholz2019cryptographic,wang2023raven}, and modern SmartNICs provide even more capabilities~\cite{kim2020case,kottur2022implementing,taranov2020srdma}.
Rewriting header fields, however, requires recomputing HMACs over the short RTP header which is feasible in programmable hardware.
Taken together, with modifications to the key-distribution protocol and implementing HMAC computation over the RTP header in hardware (as has been demonstrated feasible before), the same security as in unmodified WebRTC can be achieved in Scallop.

%% file: sec/8-related.tex
\section{Related Work}%
\label{sec:related}

\noindent
\textbf{Studies on video-conferencing applications.}
Baset and Schulzrinne's analysis of Skype, a famous P2P VoIP application in 2003~\cite{baset:analysis-of-skype}, provided crucial insights about Real-Time Communication (RTC) systems.
In addition to earlier studies comparing different VCAs~\cite{nistico:comparative-study-rtc, lee:demystifying-vca}, two recent papers conducted extensive QoE-centric measurement studies of different VCAs such as Zoom, Meet, Teams, and WebEx~\cite{chang:can-you-see-me,macmillan:perf-video-conferencing}. Important in the context of this work, these studies shed light on the infrastructure, geographic location, lag, latency, bit rate and network utilization.
Choi et al. did a more longitudinal and microscopic analysis of Zoom~\cite{choi:zoom-session-quality} while Michel et al. did an in-depth study of Zoom traffic in a production network, demystifying the Zoom RTP packet format~\cite{michel:passive-measurement-zoom}.

\noindent
\textbf{Handling video-conferencing traffic in the data plane.}
Edwards and Ciarleglio showcased a programmable data plane 
that can parse, process, and produce ``clean'' video switching of uncompressed HD video flows based on the RTP timestamp~\cite{edwards:ts-aware-rtp-switch-p4}. 
However, this demo focuses more on showing the capability of parsing RTP headers and making forwarding decisions based on RTP timestamp values with real hardware, rather than offloading any application-layer functionality (like an SFU).
More recent work showed that programmable data planes and eBPF/XDP programs running on servers can help with the NAT traversal functionality in video-conferencing applications~\cite{kfoury:offload-media-prog-switches,levai:supercharge-webrtc}. Both solutions offload the NAT traversal functionality, which is a heavy packet-processing workload, to a programmable data plane (e.g., Intel Tofino)~\cite{kfoury:offload-media-prog-switches} or an eBPF/XDP program~\cite{levai:supercharge-webrtc}. Our work goes way beyond this and enables the data plane to also perform packet replication and selective forwarding.
The closest work~\cite{pavlos:sfu-impl-p4} builds an SFU in software (bmv2) using P4 but does not actually implement a functional VCA.
Instead, it generates dummy UDP packets, using only port numbers to distinguish media layers. Forwarding relies on static multicast groups, requiring that each participant \emph{sits} behind a different physical switch port. The design omits RTP parsing and ignores all challenges related to feedback.


%% file: sec/9-conclusion.tex

\section{Conclusion}
\label{sec:conclusion}


\noindent
Taken together, our SDN-inspired SFU-switch design is driven by the key insight that most SFU tasks are, in fact, replicating and dropping media packets.
Unlike traditional infrastructure where an SFU server takes care of everything, our prototype comprises an efficient programmable data plane that processes media packets at line rate and a software control plane that handles infrequent tasks such as signaling, connection setup, quality monitoring, and rate adaptation.
Our prototype is built with a commodity server and a real programmable switch (Intel Tofino2~\cite{intel:tofino}).
Our analysis and evaluation show that Scallop delivers 7-210$\times$ improved scalability over a 32-core SFU server. 
We discuss directions for future work and other trends in VCAs today and how they related to Scallop in appendix~\ref{apx:additional-discussion}.

%% file: sec/a5-discussion.tex



\section{Discussion}
\label{apx:additional-discussion}

\boldpara{Extended VCA Use Cases.}~%
Video-conferencing applications often implement additional features that need direct media access; these include live transcription and visual effects (e.g., virtual backgrounds or ``funny hats'')~\cite{w3c:webrtc-ext-use-cases}.
In most cases, including Zoom, such functions are implemented at SFUs, and this requires decryption of client's media stream at the SFU. Because \sys{} operates on encrypted and encoded media, these features cannot be implemented in \sys{}’s data plane.
We argue, however, that these features should be implemented at clients, not SFUs, to preserve end-to-end encryption. 
WebRTC provides APIs for this~\cite{w3c:webrtc-encoded-transform, w3c:webrtc-insert-streams}.


\boldpara{SFU cascading.}~%
Selective Forwarding Unit (SFU) cascading is a technique of deploying SFUs in a hierarchical manner to improve the scalability of the infrastructure that supports video-conferencing applications by aggregation of media streams at higher-level SFUs. We argue our system is not an alternative to or a competing solution with SFU cascading. In fact, the approaches are independent and could be combined to improve the scalability further. Our control/data plane split has the potential to simplify deploying many SFU data planes under the management of a single controller. Our current system is already designed in this way and would provide the architectural framework to enable such SFU topologies.

\boldpara{SmartNIC as an alternative target.}~%
Given the stronger computational capability and flexibility, including the ability to run more advanced cryptographic functions, SmartNICs are a promising alternative target in place of programmable switches like the Intel Tofino~\cite{intel:tofino}. Of course, port density and data rate are issues for SmartNIC-based solutions as traffic and infrastructure scale up. Regardless of the target, our work presents an excellent division of labor to support large video-conferencing traffic loads, where packet processing-heavy tasks are offloaded to hardware.

\boldpara{Hardware-amenable wire protocols.}~%
The WebRTC framework and its protocols, which most video-conferencing applications follow, are not hardware-friendly. 
This is because the protocol was initially designed as a P2P protocol, where software 
with complex algorithms is designed to run on end-host devices. Furthermore, codec
and protocol designers emphasize efficiency, making every bit count and aiming to minimize the bandwidth usage between participants. This is a noble goal, but makes the protocol inherently challenging to parse and process in hardware (e.g., network devices). In particular, the complex compression algorithm and dynamic variable-length header fields make parsing and processing challenging, even with programmable data planes. 
Following the trend of hardware offloading and co-designing software and hardware, it is worth re-exploring the design and implementation of more hardware-amenable protocols, shifting the balance towards offloading computation and less on bandwidth optimization. In fact, the community has seen ideas in the past, for example using fixed-length header fields in BGP routing~\cite{kiyak:hair}.\\

%% file: sec/a2-zoom-api.tex

\section{Zoom API Data Set}
\label{apx:zoom-api}
\begin{figure}[t]
    \centering
    \includegraphics[width=0.9\columnwidth]{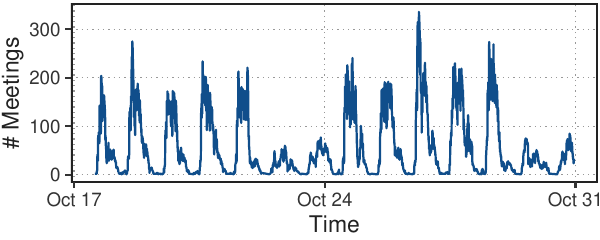}
    \caption{Number of concurrent Zoom meetings hosted by our university's Zoom account over time.}
    \label{fig:meetings-over-time}
\end{figure}

\begin{figure}[t]
    \centering
    \includegraphics[width=0.9\columnwidth]{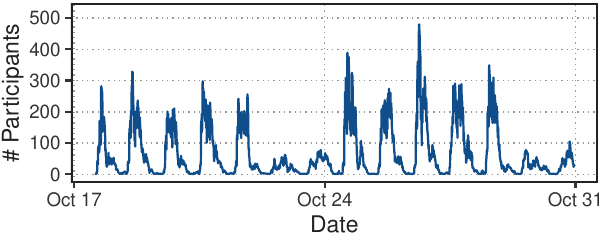}
    \caption{Number of concurrent Zoom participants hosted by our university's Zoom account over time.}
    \label{fig:participants-over-time}
\end{figure}

\noindent
The Zoom API data set used in this study was collected from our university campus.
Zoom's API~\cite{zoom-api} is made available to account administrators to access information about meetings, participants, and recordings.
We cooperated with our campus IT department to continuously collect this data which was then anonymized and aggregated to protect user privacy (see~\S\ref{sec:intro}).
The data set contains information about 19,704 meetings that took place between October 17 and October 30, 2022.
The data only includes meetings that were hosted by our university's Zoom account and does not contain any information about external meetings, for example those hosted by other institutions or individual users.
The data set includes information about the number of participants in each meeting, the duration of the meeting, and the duration and composition of media streams that were active during the meeting.
The number of concurrent meetings and concurrent participants over time are depicted in Figures~\ref{fig:meetings-over-time} and~\ref{fig:participants-over-time}, respectively.

%% file: sec/a3-zoom-passive.tex

\section{Zoom Packet-Capture Data Set}
\label{apx:zoom-passive}

\noindent
The packet-level trace we use to analyze the operation of Zoom's SFUs was collected at two border routers on our University campus on May 5th, 2022 over the course of 12 hours.
As opposed to the API data set (see Appendix~\ref{apx:zoom-api}, this data set includes packet-level data of all Zoom calls that traverse our campus network regardless of who hosted the meeting.
The key statistics of the trace are summarized in Table~\ref{tab:capture-summary}.
In order to reduce the data rate to be analyzed in software, we wrote a P4 program for an Intel Tofino switch that filters Zoom packets.
During the capture, our switch processed an average of 626,069 packets per second, with an average of 43,733 per second being Zoom traffic and, subsequently, filtered out.

\begin{table}[h]
    \small
    \begin{tabular}{ll}
        \hline
        Capture duration & 12h \\
        Zoom packets & 1,846 M (42,733/s) \\
        Zoom flows & 583,777 \\
        Zoom data & 1,203 GB (222.9 Mbit/s)\\
        RTP media streams & 59,020 \\
        \hline
    \end{tabular}
    \vspace{0.2cm}
\caption{Capture Summary}
    \label{tab:capture-summary}
\end{table}

\begin{figure}[t]
    \centering
    \includegraphics[width=0.9\columnwidth]{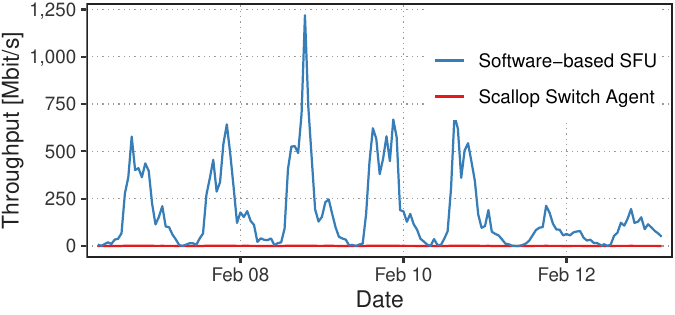} 
    \caption{Bytes processed during 2nd week of Feb '23}
    \label{fig:switch-agent-bw}
\end{figure}

\noindent
In Figure~\ref{fig:switch-agent-bw}, the blue curve shows the byte rate a software-based SFU would have to process if it were to handle all of our campus traffic during a week (peaks around 1250 Mbit/s), and the red curve shows the byte rate that the Scallop switch agent would have to process in comparison (peaks around only 4.4 Mbit/s). Even for a powerful 40 Gbit/s server, the peak byte rate from our campus already consumes 3.1\% of its total capacity if a software-based SFU is used. For large providers like Zoom, many such servers would, therefore, be required to serve the thousands of campuses and enterprises that use their services. In contrast, with Scallop, only 0.01\% of the server's capacity would be used.

%% file: sec/a4-zoom-simulcast.tex

\section{Use of Media Scalability in Zoom Packet Trace}
\label{apx:zoom-simulcast}

\begin{figure}[t]
    \centering
    \includegraphics[width=0.98\columnwidth]{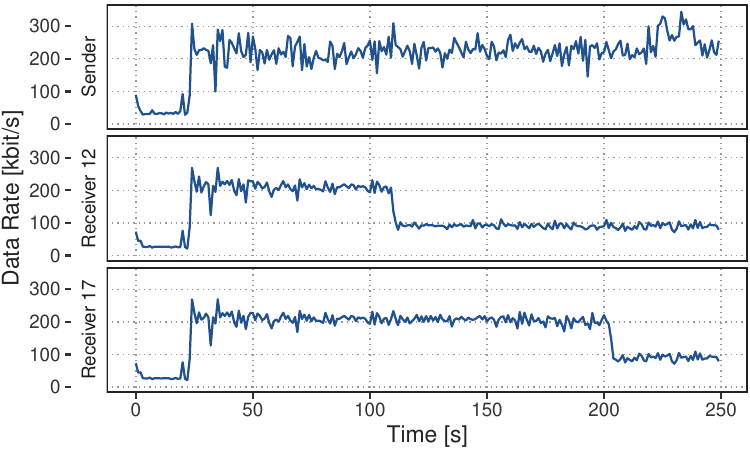} 
    \caption{Forwarded bytes of a single video stream to two separate participants in a Zoom meeting.}
    \label{fig:kbits-per-stream}
\end{figure}

\begin{figure}[t]
    \centering
    \includegraphics[width=0.98\columnwidth]{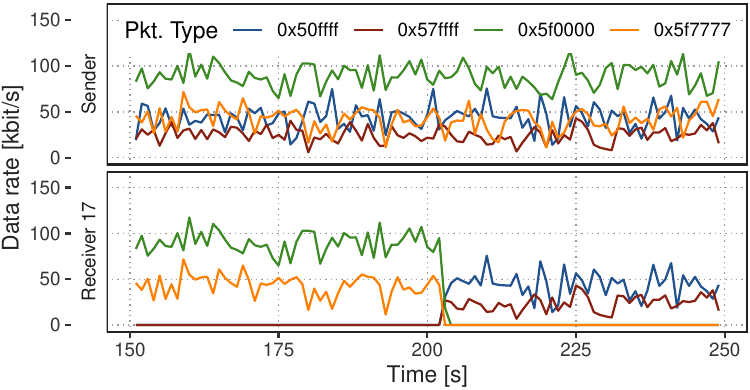} 
    \caption{Forwarded bytes per scalability layer of a single video stream to a single participant in a Zoom meeting.}
    \label{fig:kbits-per-layer}
\end{figure}

\begin{figure}[t]
    \centering
    \includegraphics[width=0.85\columnwidth]{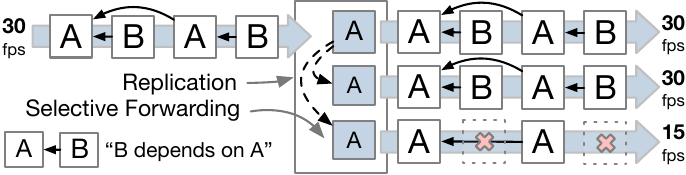}
    \caption{Media forwarding when using SVC.}
    \label{fig:sfu-operation}
\end{figure}

\noindent
Scalable Video Coding (SVC) enables efficient video transmission by encoding a stream into multiple layers of increasing quality, allowing selective adaptation based on network conditions. An SFU can reduce the frame rate (or resolution) by dropping frames from the enhancement layers, effectively limiting playback to a lower frame rate while maintaining smooth video delivery. Figure~\ref{fig:sfu-operation} shows an example where, by discarding every other frame, the SFU decreases the temporal resolution, reducing bandwidth usage while preventing playback freezes.

To give a concrete example of how Zoom adapts media streams at their SFUs, we analyze a meeting that is part of our Zoom packet trace introduced in Appendix~\ref{apx:zoom-passive}.
Figure~\ref{fig:kbits-per-stream} shows an example of a participant's outgoing video stream and the corresponding incoming streams at two other participants.
Time zero denotes the start of the meeting at which the sending participant starts transmitting a low-bitrate video stream which then increases in bitrate at around 20 seconds into the meeting.
Presumably in response to constrained downlink capacity, the SFU then reduces the bitrate forwarded to participant 12 (at around 110s) and the bitrate forwarded to participant 17 (at around 200s), respectively.

Zoom uses the Real-time Transport Protocol (RTP)~\cite{rtp-rfc} encapsulated in custom, proprietary headers over UDP to transmit media~\cite{michel:passive-measurement-zoom}.
The RTP header carries a sequence number and various extension fields in each packet that is not being changed by Zoom's servers during forwarding.
As a result, we can exactly specify which packets are forwarded to a given participant by the SFU and which are not.
We further observed that Zoom's RTP packets carry various RTP extension headers.
One header extension field carries a three-byte value that appears to be a bit mask.
The packets in the outgoing media streams typically carry between two and six different values for this field across their packets which we for now denote as the \emph{packet type}.
We observe that the SFU only forwards either all or a strict subset of these packet types to a given 

Figure~\ref{fig:kbits-per-layer} shows the bit rate of the same media stream as previously shown broken down by packet type.
We can see that the adaptation of the media stream received by participant 17 around second 200 is achieved by changing the set of packet types forwarded to this client.
This is consistent with previous observations that Zoom uses media scalability through Scalable Video Coding (SVC) and Simulcast~\cite{michel:passive-measurement-zoom,how-zoom-works, w3c:webrtc-svc, schwarz:svc-overview} and leverages this field to indicate to the SFU which layer of a scalable stream the respective packet carries.
It is common to encode the type of scalability layer in the RTP header~\cite{schwarz:svc-overview} which is transmitted in clear text; 
similar header extensions exist, for example, for the AV1 codec~\cite{av1, av1-rtp} which we leverage in this work.

%% file: sec/a6-rtp-extension-parsing.tex

\section{RTP Extension Parsing}
\label{apx:rtp-extension-parsing}

\noindent
After parsing the Ethernet, IP, and UDP headers, we use the destination IP and port numbers to determine whether the packet is destined for the SFU. If so, we \emph{lookahead} into the first $4$ bits of the UDP payload to determine whether the packet resembles an RTP or an RTCP packet. If the packet is an RTP media packet or an RTCP sender report, we mark it for replication; otherwise, we forward it.
Rate adaptation in SVC works by dropping packets that correspond to the higher quality layers of an AV1 media stream, as needed. Determining what quality layer a packet belongs to requires parsing into the RTP \emph{header extensions} beyond the RTP header, identifying the header extension corresponding to the AV1 protocol (if present), and parsing the AV1 header\footnote{We implement a Wireshark plugin to parse RTP header extensions to identify and display AV1 headers which we will open-source, along with all of our other source code, pertaining to both the control and data planes.} to extract its \emph{dependency template ID}, which when combined with the control plane's knowledge of the mapping between dependency templates and quality layers, tells us the quality layer of the packet itself. Parsing deep into the RTP header extensions to find the AV1 header requires handling variable-length and variable-position headers which is tricky because the P4 parser on Tofino relies on a largely static parse graph. We solve this by (i) implementing a \emph{depth-aware} parsing tree for the RTP header extensions, where for every depth in the tree, we have a \emph{landing state} that determines---using the \emph{lookahead} function---what type of header element (one-byte header, two-byte header, or padding) comes next, and (2) using the Tofino's \emph{ParserCounter} feature to track whether there are header bytes left to parse.